\documentclass[11pt]{article}
\renewcommand{\baselinestretch}{1.1}
\usepackage{fullpage}
\usepackage{hyperref}
\usepackage{amssymb,amsfonts,amsmath}
\usepackage{helvet}

\usepackage{amsmath}
\usepackage{amssymb}

\usepackage{graphicx}

\usepackage{cite}

\usepackage{color} 
\usepackage{enumerate}
\usepackage{color,soul}
\newcommand{\blank}[1]{}
\newcommand{\lt}{\left}
\newcommand{\rt}{\right}
\newcommand{\la}{\langle}
\newcommand{\ra}{\rangle}
\newcommand{\nn}{\nonumber}

\newcommand{\nuff}{\nu_{\textrm{eff}}}
\newcommand{\phin}{\phi_{\textrm{in}}}
\newcommand{\phout}{\phi_{\textrm{out}}}
\newcommand{\Phin}{\Phi_{\textrm{in}}}
\newcommand{\Phout}{\Phi_{\textrm{out}}}

\blank{\usepackage{lineno}
\linenumbers
\newcommand*\patchAmsMathEnvironmentForLineno[1]{%
  \expandafter\let\csname old#1\expandafter\endcsname\csname #1\endcsname
  \expandafter\let\csname oldend#1\expandafter\endcsname\csname end#1\endcsname
  \renewenvironment{#1}%
     {\linenomath\csname old#1\endcsname}%
     {\csname oldend#1\endcsname\endlinenomath}}%
\newcommand*\patchBothAmsMathEnvironmentsForLineno[1]{%
  \patchAmsMathEnvironmentForLineno{#1}%
  \patchAmsMathEnvironmentForLineno{#1*}}%
\AtBeginDocument{%
\patchBothAmsMathEnvironmentsForLineno{equation}%
\patchBothAmsMathEnvironmentsForLineno{align}%
\patchBothAmsMathEnvironmentsForLineno{flalign}%
\patchBothAmsMathEnvironmentsForLineno{alignat}%
\patchBothAmsMathEnvironmentsForLineno{gather}%
\patchBothAmsMathEnvironmentsForLineno{multline}%
}
}

\begin{document}
\begin{flushleft}
{\Large
\textbf{Cross-scale neutral ecology and the maintenance of biodiversity}
}
\\
James P. O'Dwyer$^1$
Stephen J. Cornell$^2$\\
\bf{1} Department of Plant Biology, University of Illinois, Urbana IL USA \\
\bf{2} Institute of Integrative Biology, University of Liverpool, Liverpool L69 7ZB, UK\\
\end{flushleft}

\textbf{Correspondence to be sent to:} \\Dr James P. O'Dwyer\\ Department of Plant Biology\\University of Illinois, Urbana IL 61801 \\ jodwyer@illinois.edu
\renewcommand{\baselinestretch}{1.1}
\normalsize



\section*{Abstract}
One of the first successes of neutral ecology was to predict realistically-broad distributions of rare and abundant species.  However, it has remained an outstanding theoretical challenge to  describe how this distribution of abundances changes with spatial scale, and this gap has hampered attempts  to use observed species abundances as a way to quantify what non-neutral processes are needed to fully explain observed patterns. To address this, we introduce a new formulation of spatial neutral biodiversity theory and derive analytical predictions for the way abundance distributions change with scale. For tropical forest data where neutrality has been extensively tested before now, we apply this approach and identify an incompatibility between neutral fits at regional and local scales. We use this approach derive a sharp quantification of what remains to be explained by non-neutral processes at the local scale, setting a quantitative target for more general models for the maintenance of biodiversity.
\pagebreak

\section*{Introduction}

Neutral biodiversity theory has become one of the most tested paradigms of macroecology ~\cite{Hubbell2001,Rosindell2011,odwyer2013eob,azaele2016statistical}.  It combines the ecological mechanisms of birth, death, competition, speciation, and spatial dispersal to make predictions for ecological patterns, and makes manifest the belief that the many differences between species may not be critical for successfully predicting large-scale, aggregated phenomena.  Subsequent studies have expanded on the original neutral approach~\cite{Etienne2005,Etienne2009},  generalizing the theory to include life history~\cite{Odwyer2009,xiao2015comparing}, fitness differences~\cite{Haegeman2011,odwyer2014redqueen,kessler2014neutral} and multiple modes of speciation~\cite{rosindell2010protracted,etienne2011neutral}.  But at the core of this theory there is  a missing link: we lack a complete picture of how neutral predictions change with spatial scale.
\smallskip

At the largest, continental scales,  drift in population sizes arising from neutral demographic processes is capable of generating a broad range of species abundances following the log series distribution~\cite{fisher1943relation}. This is a classic and plausible distribution,  fitted to many data sets~\cite{white2012characterizing}, but hard to measure directly due to the huge scales involved. Local community data, collected at the scale of hectares, has provided a more tractable way to test neutral predictions. At these scales we would expect dispersal limitation to significantly affect the distribution of species abundances alongside birth, death, and competition---inevitably, not every species will have the same abundance in every location. But existing neutral predictions for local community abundances are spatially-implicit, meaning that local community data must be fitted using two effective parameters that characterize the input from the surrounding region.  These parameters are difficult to interpret in terms of biological processes that could be verified independently, or used to make predictions at different spatial scales from the data set used to fit the model. The result is often a successful description of local community abundances. But the freedom to fit these parameters means that we may be obtaining the right species abundance curve,  for the wrong reasons.

\smallskip

Spatially explicit neutral theory overcomes this problem by modelling dispersal with a dispersal kernel, corresponding to a process that can be verified and interpreted independently.  Progress towards building a spatially explicit model of neutral biodiversity has taken multiple forms, and each has some benefits and drawbacks. These include numerical simulations~\cite{Hubbell2001,Rosindell2008,Rosindell2009,rosindell2013universal}, which can become unfeasible for very small speciation rates and very large systems; hybrid approaches where non-spatial parameters are fitted to a spatially-explicit simulations~\cite{Chisholm2009,etienne2011spatial}; the limit of very short-scale dispersal~\cite{durrett1996spatial}; a focus on predicting pairwise correlations in species composition, but not species abundances~\cite{Chave2002b,houchmandzadeh2003clustering}; phenomenological models~\cite{peruzzo2016phenomenological}; and analytical approaches that make statistical assumptions which are violated in real communities~\cite{Odwyer2010,grilli2012absence}. The studies have shown that spatially explicit neutral models predict cross-scale patterns of species abundance that resemble empirical patterns qualitatively~\cite{rosindell2013universal}, but differ in detail from the predictions of the original, spatially implicit theory~\cite{etienne2011spatial}.  However, while the spatially explicit theory makes more realistic assumptions, only the spatially implicit theory has been compared exhaustively to empirical abundance patterns. The prime reason is that analytical methods for computing these abundance distributions have, until now, only been available for the spatially implicit theory.

\smallskip

In this paper, we address this  gap by introducing a new mathematical formulation of the spatial theory of neutral biodiversity, derived using the backward equation formulation of stochastic processes.  While an exact solution of these equations is not available due to non-linearities, we have identified an accurate approximation scheme which we test extensively using spatially-explicit numerical simulations.   These new results allow us to connect local observations and large-scale data.  We subsequently parametrize the neutral model using sparse, regional and continental-scale observations, and go on to test whether it is then consistent with distributions of abundance at the local scale.  We focus on data that has already been fitted using spatially-implicit models, to see whether our spatially-explicit approach deviates from these earlier results. 

\smallskip

Combining our modeling approach with data from these multiple scales, we find that neutrality alone significantly underestimates local species diversity, and also deviates from the observed distribution of rare and abundant species.  Our intuition might have been that dispersal and neutrality would lead to many rare, transient species, which disperse into a local community and quickly drop out before proliferating.  In fact, our spatial neutral prediction dramatically under-predicts the observed number of rare species. This indicates that local stabilizing mechanisms are likely important to understand and accurately predict local patterns of biodiversity~\cite{Mangan2010,Comita2010,chisholm2011theoretical}, and precisely quantifies what remains for these approaches to explain.

\section*{Results}

Our model is based around the neutral assumptions of intrinsic birth and mortality rates that are identical across all species, in addition to symmetric competition for a single resource, which we approximate using the mean field approach~\cite{odwyer2014redqueen}. This is also known as a non-zero sum formulation~\cite{Etienne2007} because the total community size is allowed to fluctuate around an average value.  The resulting model is an assemblage of ecologically-identical species, with a constant, total density across space and time when in steady state.  New species enter the community via speciation, which occurs at a fixed per capita rate, and hence a fixed rate per unit time and area. All species eventually leave the community due to extinction.  So the model reduces to a set of independent populations, beginning their existence with a single individual, and proliferating transiently across space. Meanwhile, we would like to predict the probability that a focal species has a given number of individuals in our sample location in the present day.  In our Supplementary Information we derive the following backward equation (so-called because we look `backwards' from the present day, as explained in our Supplementary materials) to characterize these dynamics and this observable:
\begin{align}
\frac{\partial P(k,A,x,y,t)}{\partial t} & =b\sigma^2\lt[\frac{\partial^2 P(k,A,x,y,t)}{\partial x^2} + \frac{\partial^2 P(k,A,x,y,t)}{\partial y^2}\rt]\nn\\& +(b-\nu)\sum_m P(k-m,A,x,y,t) P(m,A,x,y,t) \nn\\
& -\left(2b-\nu\right)P(k,A,x,y,t)+b\delta_{k,0}.\label{eq:back}
\end{align}
In this equation, $x$ and $y$ represent the location of the focal species' initial individual in two-dimensional space, while $t$ is how long ago from the present day this species entered the community, and $A$ is the sample area in the present day. $P(k,A,x,y,t)$ is then the probability that a species with initial location given by coordinates $x$ and $y$ has $k$ conspecifics in the sample region after time $t$.  What processes determine this observable? $b$ is the intrinsic birth rate, while $\nu$ is the per capita speciation rate. $\sigma$ characterizes the spatial process, and can be thought of as proportional to the root of the mean squared distance that a seed is dispersed. 

\smallskip

\begin{figure}
\includegraphics[angle=0,scale=0.45,clip=FALSE,trim= -0 0 0 0 ]{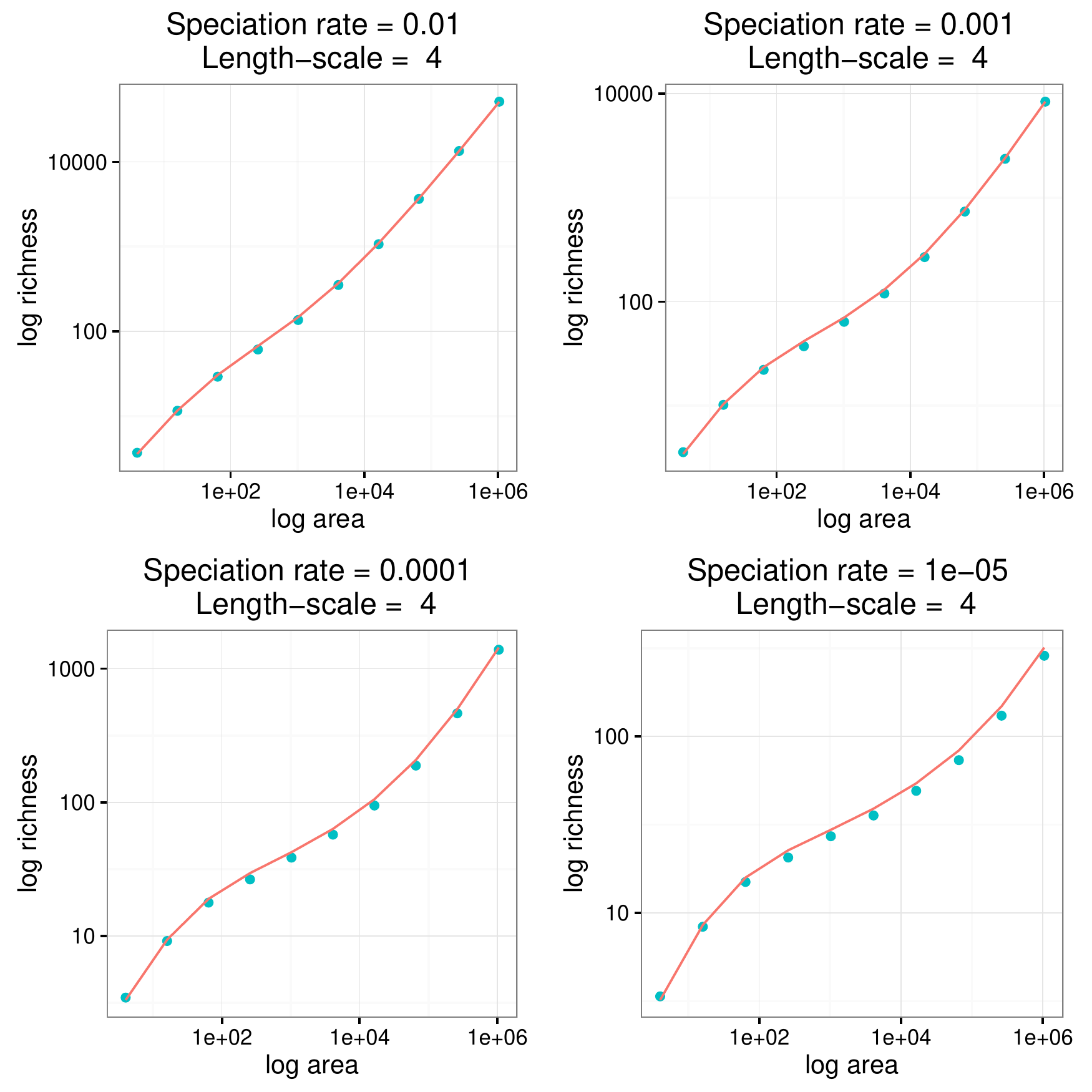}
\includegraphics[angle=0,scale=0.45,clip=FALSE,trim= -0 0 0 0 ]{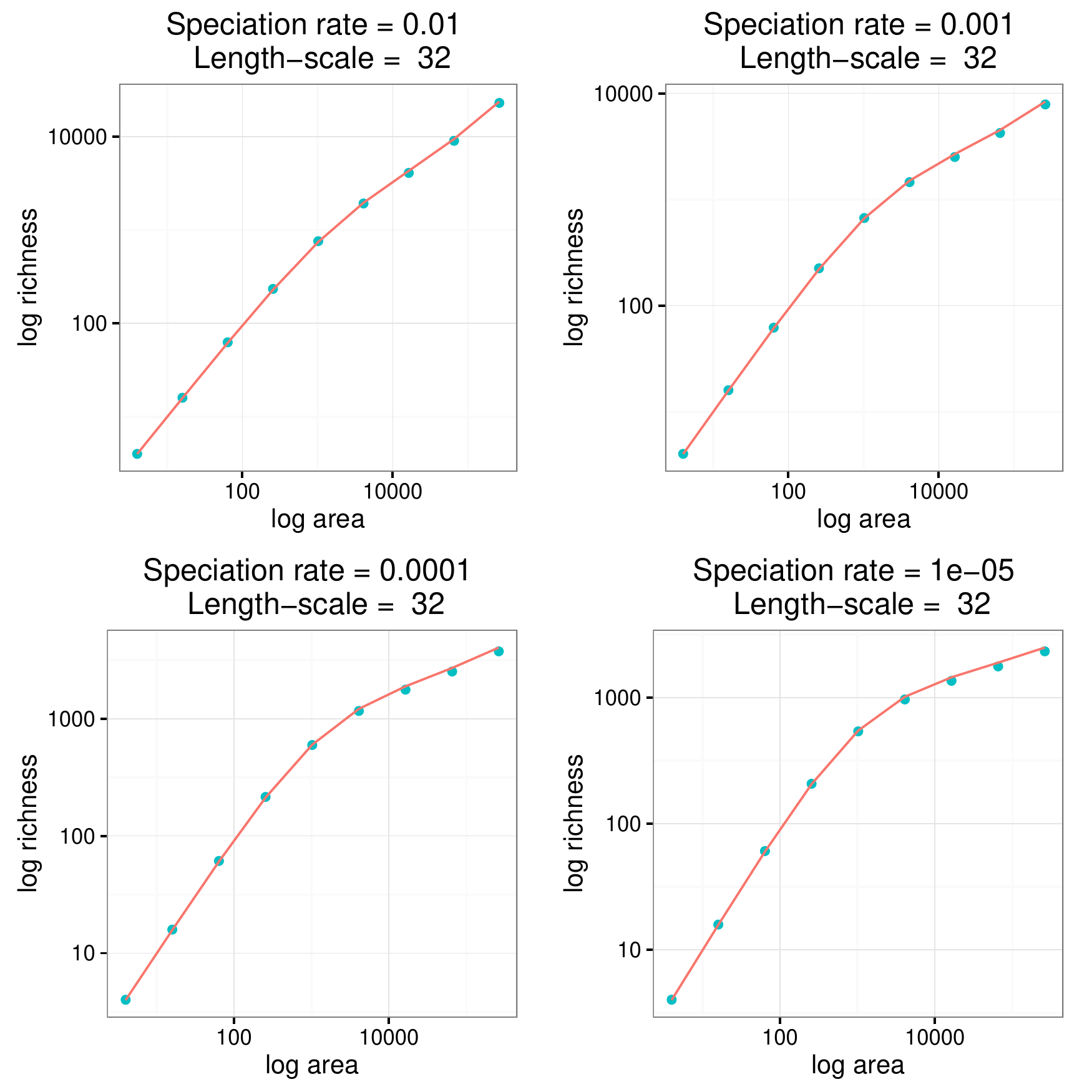}
\put(0,100){\includegraphics[angle=0,scale=0.5,clip=FALSE,trim= -0 0 0 0 ]{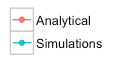}}
\caption{\textbf{The species-area curve}. We show a comparison between species richness as a function of sampled area for our analytical approximation to spatial neutral theory, compared with numerical simulations~\cite{Rosindell2008}.  Over this range of values of speciation rate ($\frac{\nu}{b}$) and dispersal length-scale ($\sigma$ in the main text), we see quantitative agreement between our approximation method and these earlier numerical results.}\label{fig:sar}
\end{figure}

The initial condition is simply that at $t=0$, if the location $(x,y)$ is inside the sample area $A$, then $P(k,A,x,y,0)=\delta_{k,1}$. Conversely, if $(x,y)$ is outside the sample area, $P(k,A,x,y,0)=\delta_{k,0}$.  Assuming for the time being that we can solve Eq.~\eqref{eq:back} for $P(k,A,x,y,t)$, then we immediately have a community level prediction for the average number of species with exactly abundance $k$ in a sample area $A$:
\begin{equation}
S(k,A) = \nu\rho\int_{-\infty}^{\infty} dx\int_{-\infty}^{\infty}dy \int_0^\infty dt\ P(k,A,x,y,t)\label{eq:sad1},
\end{equation}
where $\rho$ is the constant average total density across space. However, solving~Eq.~\eqref{eq:back} with the appropriate initial condition is non-trivial, due to the quadratic terms in $P$, which derive from the birth process, and we do not know of any closed-form solution.  This non-linearity is the essence of why this is is a difficult problem, and is also reflected in the challenge of finding exact solutions in the corresponding forward-in-time, field theory version of this model~\cite{Odwyer2010}. 

\smallskip

\subsection*{Species-area curve} In our Supplementary Information we introduce an approximation scheme to linearize Eq.~\eqref{eq:back}, with different linearizations applying in different regions of the landscape.  As a special case of Eq.~\eqref{eq:sad1}, we first focus on solutions for the Species-area curve, which counts the total number of distinct species (with any value of $k>0$) as $A$ increases, in this case for a circular sample region. We find the following approximate solution for this relationship: 
\begin{align}
S(A) & =  \rho\nuff A \nn\\
& +\frac{2\rho\sqrt{A\pi\sigma^2}\lt(1-\nuff\rt)I_1(\frac{\sqrt{A}}{\sqrt{\pi\sigma^2}})}{\frac{1}{\sqrt{\nuff}}I_1\lt(\frac{\sqrt{A}}{\sqrt{\pi\sigma^2}}\rt)\frac{K_0\lt(\sqrt{A\nuff/\pi\sigma^2}\rt)}{K_1\lt(\sqrt{A\nuff/\pi\sigma^2}\rt)}+I_0\lt(\frac{\sqrt{A}}{\sqrt{\pi\sigma^2}}\rt)}\label{eq:sar}
\end{align}
In this solution, we have used the short-hand  $\nuff= \frac{\nu}{b-\nu}\log(b/\nu)$, but no new parameters have been introduced, while $I_n$ and $K_n$ are modified Bessel functions. Note that only the per capita, per generation speciation rate, $\nu/b$ enters this solution, and so the rates $b$ and $\nu$ do not independently affect the Species-area curve. How well does this approximation work?  In Fig.~1 we demonstrate the agreement between theoretical and simulated curves over a range of speciation rates and values of $\sigma$.

At small areas, with $A<<\pi \sigma^2 $, both simulations and theoretical results give $S(A)\simeq \rho A$, i.e. where most new individuals belong to distinct species as the sample area is increased.  At large areas, $A>>\pi\sigma^2/\nuff$, both simulations and theoretical results approach  $S(A)\simeq \rho \nuff A $, so that richness again increases linearly with area, but with a smaller overall coefficient.  In between these extremes, we also see good agreement between the simulated and theoretical curves.   The transition between large and intermediate scales has been modeled before, by making various phenomenological assumptions about species range shapes and distributions~\cite{allen2003effects,Storch2012}.  Here we can see that explicitly the first  correction to large-scale linear behavior is proportional to $\sqrt{A}$, identical to these earlier results~\cite{allen2003effects}, so that at intermediate to large scales:
\begin{equation}
S(A)\simeq \rho \nuff A+\frac{2\rho\sqrt{A\nuff\pi\sigma^2}}{\sqrt{\nuff}+1},
\end{equation}
again only valid when $A\nuff/\pi\sigma^2>>1$. This agreement is non-trivial, given that the shape of any given neutral species range will not satisfy the simplifying assumptions (of circularity or convexity) made in the phenomenological approaches. Finally, the intermediate region as a whole has been fitted to empirical data drawn from across many taxa and enviroments using a power law~\cite{Arrhenius1921}, and our resuls show that in neutral theory the power law SAC can only ever be an approximate description.

\subsection*{Spatial Scaling of the Species Abundance Distribution}

We now apply the same approximation method to solve for the species abundance distribution, $S(k,A)$, given by Eq.~\eqref{eq:sad1}. Our solution is expressed in terms of the generating function, $\Psi(z,A) = \sum_{k=1}^{\infty} S(k,A) z^k$, and in our Supplementary Information we supply R code to quickly (and with quantifiable error) extract the SAD itself from this generating function, following the method of~\cite{bornemann2011accuracy}. Our solution for this generating function is given by:
\begin{align}
\Psi(z,A) & = S(A)-\rho f(z) A \nn\\
& +\frac{ \frac{2(f(z)-(1-z))}{\sqrt{h(z)}}\rho\sqrt{A\pi\sigma^2}\ I_1\lt(\sqrt{\frac{h(z)A}{\pi\sigma^2}}\rt)}{I_0\lt(\sqrt{\frac{Ah(z)}{\pi\sigma^2}}\rt)+\sqrt{\frac{1-z}{f(z)}}\frac{K_0\lt(\sqrt{\frac{Af(z)h(z)}{(1-z)\pi\sigma^2}}\rt)}{K_1\lt(\sqrt{\frac{Af(z)h(z)}{(1-z)\pi\sigma^2}}\rt)}I_1\lt(\sqrt{\frac{Ah(z)}{\pi\sigma^2}}\rt)}\label{eq:sad}
\end{align}
where we have defined the functions $f(z) = \frac{\nu}{b-\nu}\log\lt[\frac{b-(b-\nu)z}{\nu}\rt]$ and $h(z) = 1-z(1-\nu/b)$ for ease of notation, and $S(A)$ is given by Eq.~\eqref{eq:sar}.  While finding the Species-area curve is already a promising step, matching the full species abundance distribution as a function of area is a much sterner test for our approximation scheme.  In Fig.~2, we show that our solution closely matches numerical simulations over a range of speciation rates $\nu$, values of dispersal length-scale, $\sigma$, and sample areas. 

This expression for $\Psi$ displays the properties of species abundance distributions that have previously been  found by simulations of spatial neutral models \cite{rosindell2013universal}.  First, when $A$ is very large, the third term becomes much smaller than the second term, so the generating function is approximately $\rho f(z) A$.  Expanding in powers of $z$, we find in this limit
\begin{align}
S(k,A)\propto \frac{\left(1-\frac{\nu}{b}\right)^n}{n},
\end{align}
which is a Fisher logseries with diversity parameter $\alpha=1-\frac{\nu}{b}$.

Second, the species-abundance distributions display the ``universality'' noted by Rosindell and Cornell \cite{rosindell2013universal}.  While the expression for $\Psi$ depends on all four quantities $z$, $A$, $\nu/b$, and $\sigma$, in Appendix 2.3 we show in that, when the speciation rate is small ($\nu/b\to 0$), it reduces to an expression that depends only on the two  combinations $Z=(1-z)b/\nu$ and $Y=A\nu/(b\sigma^2)$.  We also show in Appendix 2.3 that this is not limited to our approximation, but is also a property of the exact solution to the backward equation. We further show in Appendix 2.3 that this is equivalent to the species abundance distribution taking the scaling form $S(k,A)=\nu\tilde{S}(k\nu,A\nu/\sigma^2)$.  This confirms analytically that species abundance distributions for spatial neutral models form a single-parameter family of curves, which extends the universality described by Storch et al~\cite{Storch2012} for species-area curves and endemics-area curves.

\begin{figure}
\includegraphics[angle=0,scale=0.45,clip=FALSE,trim= -0 0 0 0 ]{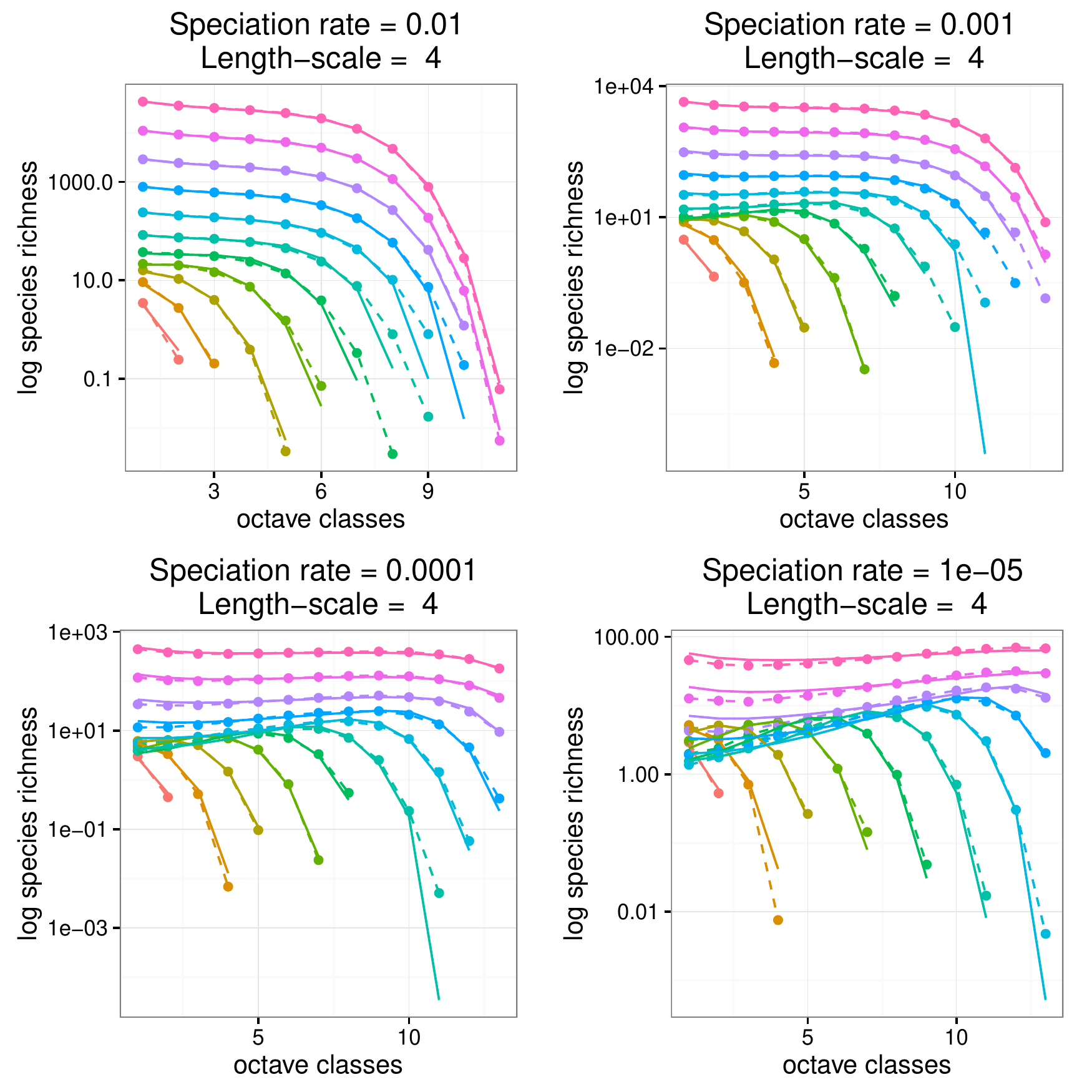}
\includegraphics[angle=0,scale=0.45,clip=FALSE,trim= -0 0 0 0 ]{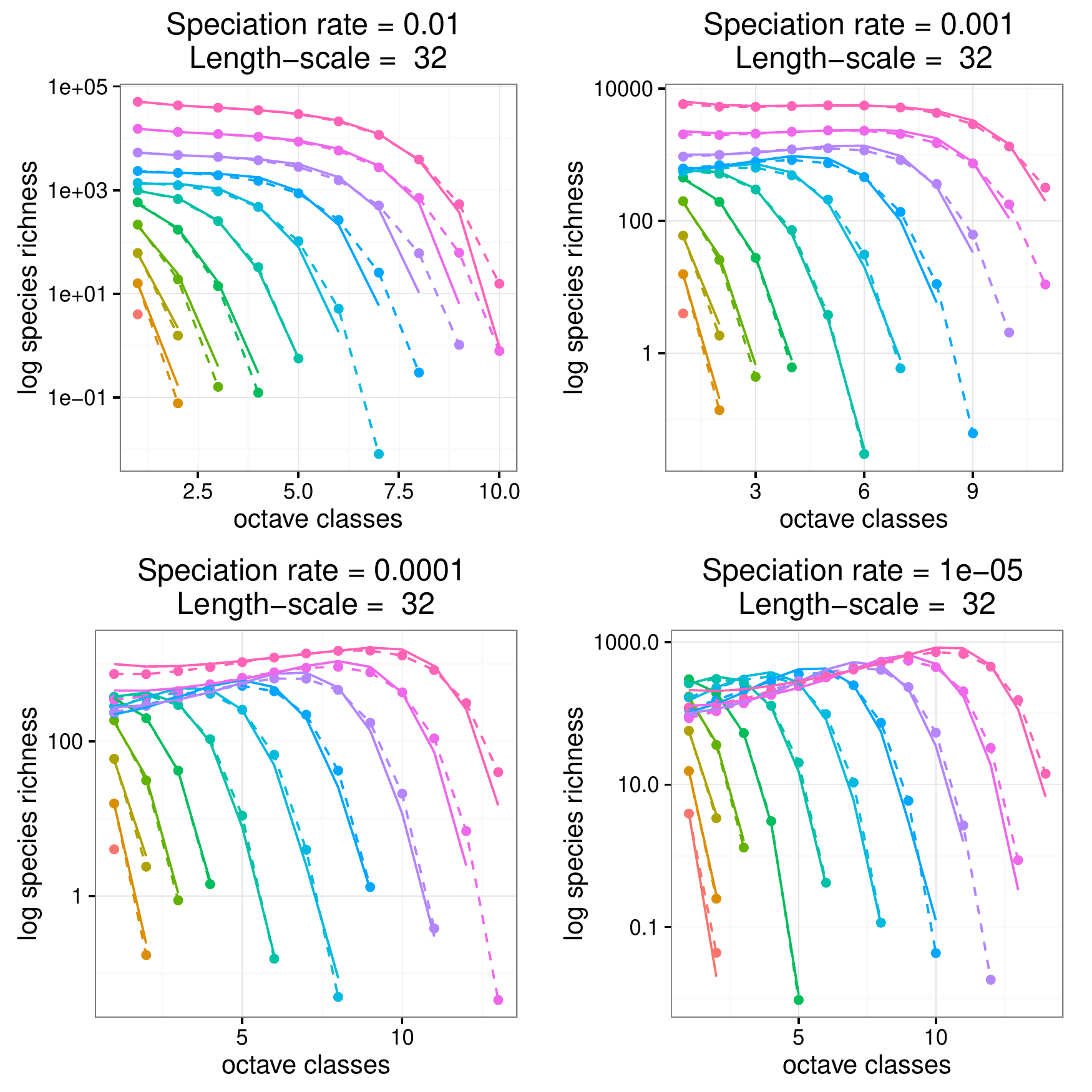}
\put(10,50){\includegraphics[angle=0,scale=0.4,clip=FALSE,trim= -0 0 0 0 ]{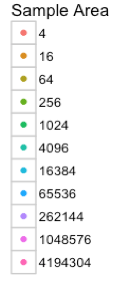}}
\caption{\textbf{The species-abundance distribution}. We test our approximation over a range of speciation rates, $\nu/b$ (and two different dispersal length-scale, $\sigma$), by comparing the predictions using Eq.~\eqref{eq:sad} with numerical simulations~\cite{rosindell2013universal}. Our results show good (though not perfect, due to both our approximations and the details of the numerical simulation) agreement over this range of parameter values. }\label{fig:sad}
\end{figure}

\section*{Application to Tropical Forest Communities}

\smallskip

Now armed with a spatially-explicit prediction for the species abundance distribution, we test whether the observed distribution of tree species abundances at the Barro Colorado Island 50ha plot (BCI) is consistent with a neutral model where parameters are fixed independently of the plot-scale counts.  Due to its high diversity and regular and comprehensive census, this plot has often been a testing ground for theoretical explanations of biodiversity patterns. It has also been extensively compared to the spatially-implicit neutral predictions, which have closely matched the observed abundance distribution~\cite{Volkov2003,Chave2004,Volkov2007}, although even early on it was emphasized that it may be difficult to distinguish neutral fits from alternatives with the same number of parameters~\cite{mcgill2003test,May1975}.  Taking our alternative route, how should we determine the parameters of our spatially-explicit model? Density $\rho$ is straightforward to estimate, and  we could conceivably match the dispersal length-scale $\sigma$ using inverse modeling and seed-trap data~\cite{clark1999seed}.  However, the speciation rate $\nu$ would be extremely challenging to measure directly, even to the extent that it is well-defined~\cite{rosindell2010protracted,etienne2011neutral}.  

\begin{figure}
\includegraphics[angle=0,scale=1.0,clip=FALSE,trim= -0 0 0 0 ]{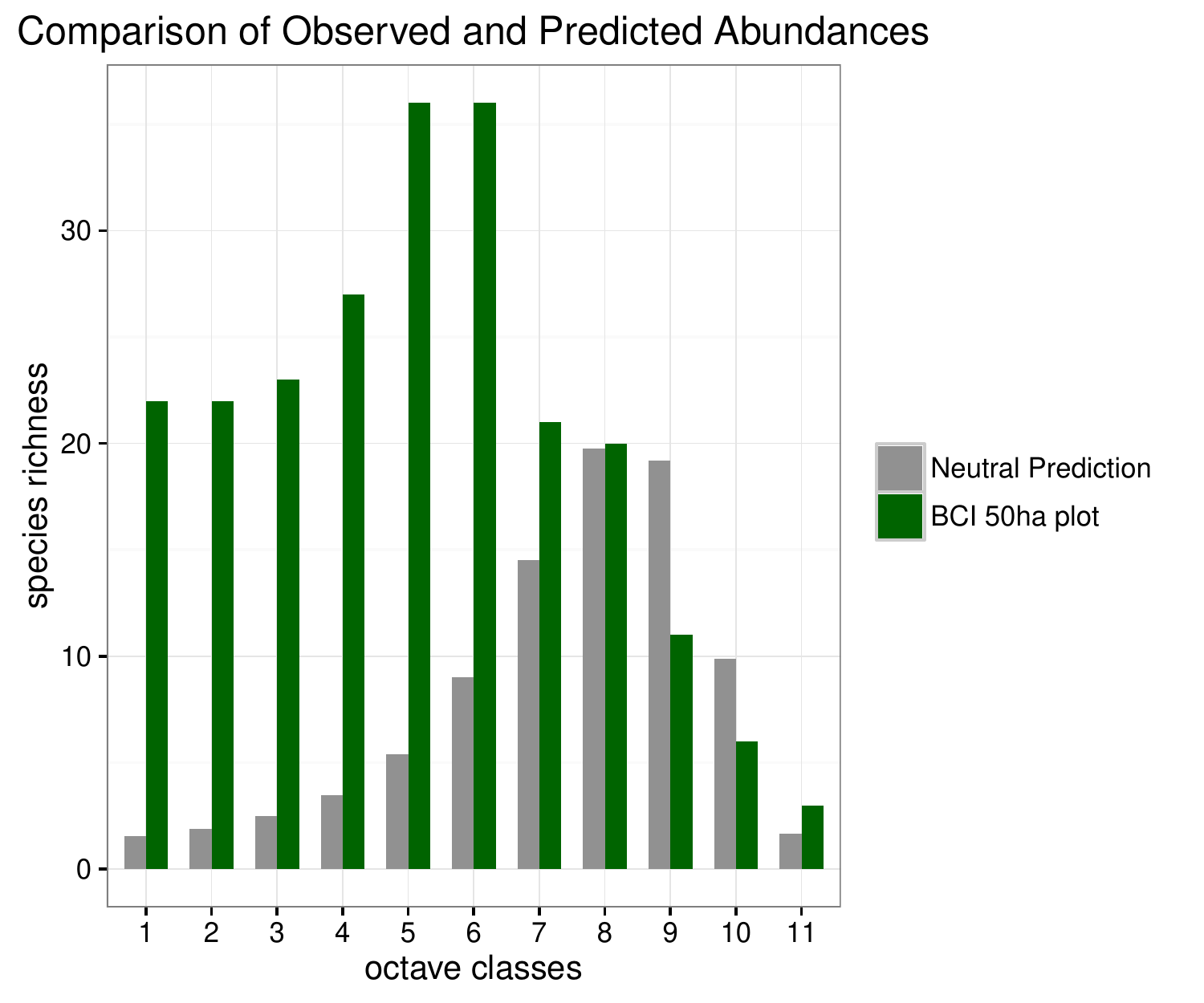}
\caption{\textbf{Neutral predictions at BCI}. This comparison demonstrates the discrepancies between neutral predictions and the observed data at the 50ha plot on Barro Colorado Island~\cite{Hubbell2005,Condit1998,Hubbell1999}. Neutral predictions are generated by fitting our spatial neutral model using large-scale data reported and analyzed in~\cite{Condit2002}. The results show that these large-scale fits produce a local-scale prediction for species abundances that both underestimates local species richness, compared with observed data, and also skews abundances from rare to more abundant species.}\label{fig:sar}
\end{figure}

\smallskip

Here we take a different approach, leveraging the methods and results of earlier studies focusing on large-scale spatial correlation functions~\cite{Chave2002b,Condit2002}. These papers focus on the two-point spatial correlation function, known as $F(r)$, the probability that two trees sampled at a separation $r$ from each other are conspecifics.  For spatial neutral theory, it has already been shown~\cite{Chave2002b,Odwyer2010} that this function takes the following form at large spatial separations:
\begin{align}
F(r) = \frac{1}{\rho\pi\sigma^2}K_0\lt(\frac{r\sqrt{2\nu/b}}{\sigma}\rt).
\end{align}
Using this result, and data from trees with diameter $>10$cm in $34$ $1$ha plots in Panama (separated by values of $r$ between $\sim 0.5$ and $\sim 50$ km), Condit et al~\cite{Condit2002} obtained parameter fits of $\sigma=40.2m$ and $\nu/b = 5.10^{-8}$. These fitted values used the observed density of $\rho\simeq 0.04 m^{-2}$.  While speciation would be difficult to estimate independently of this fit, this value of $\sigma$ is similar to those obtained from seed-trap data~\cite{clark1999seed}.

\smallskip

With these parameters fixed, we can test whether these large-scale data are consistent with the observed distribution of species abundance at the $50$ha plot scale~\cite{Hubbell2005,Condit1998,Hubbell1999}. Fig.~3 demonstrates that the spatial neutral model severely underestimates diversity at the $50$ha scale, by approximately a factor of two.   It also  skews the distribution of species abundances towards more dominant species, with singleton species (those with just one stem $>10$cm in the plot) underestimated by a factor of around twenty compared with observed counts.   We are only looking at one plot, but this is a data set where spatially-implicit neutral predictions had already passed a series of tests, and so it is important to see whether these hold up when the neutral model is spatially-explicit.  In summary, the formulation of spatial neutral theory we have considered here allows us to show that local abundances are not consistent with the parameters inferred from large-scale data. 

\smallskip

\section{Discussion}

Neutral theory has most often been formulated in a spatially-implicit way, so that local species abundance distributions depend on two free parameters characterizing the influx from a larger (but unmeasured) regional community~\cite{Hubbell2001,Volkov2003}.  These parameters can be roughly thought of as determining the richness of this larger community and then rate of immigration from the regional to the local scale. It is certainly difficult to estimate the richness of this larger community, and while some model approaches have attempted to connect the immigration rate to explicit mechanisms of dispersal~\cite{Chisholm2009,chisholm2012linking}, this matching only works in certain idealized limits. It has therefore been difficult to know the values of these two fitted parameters are biologically reasonable or not, even when the neutral theory successfully matches the distribution of species abundances in a local community. 

\smallskip

We have introduced a new formulation of spatially-explicit, stochastic biodiversity theory that complements and extends the predictions of earlier approaches~\cite{Chave2002b,Odwyer2010,Rosindell2008}.  Making predictions from our model reduces to the solution of a non-linear partial differential equation, and while it is unlikely that this equation has a closed form solution, we identified an approximation scheme which closely matches the quantitative results of numerical simulations. We focused on predictions for the species-area curve, and for the distribution of species abundances as a function of spatial scale.  The latter prediction is a key advance over earlier formulations of neutral theory, as it allows us to test whether neutral theory matches observed abundance distributions without tuning parameters to fit this data.

\smallskip

It is uncontroversial to say that neutrality is an incomplete description of any given natural system. Instead, neutrality provides a starting point from which we might hope to infer the importance of non-neutral processes.  The species abundance distribution has been largely written off as an approach to achieving this, in part because spatially-implicit neutral models are flexible enough to fit a vast range of different local abundance distributions.   In applying our spatially-explicit methodology to Panamanian tropical forest data, we in part rehabilitate the species abundance distribution as a diagnostic for what is missing from the neutral explanation, in an approach consistent with previous calls to test multiple patterns simultaneously, rather than just species abundances alone~\cite{mcgill2007species}.  Specifically, we identified a mismatch between large-scale pairwise correlation data, local community abundances, and neutrality: by fitting neutral parameters using large-scale data for the pairwise-similarity of widely separated plots, we were able to show that the corresponding neutral prediction for species abundances  underestimates diversity at the 50ha scale, and dramatically skews the distribution of abundances away from rare species. This shows that Gaussian dispersal limitation alone is unlikely to explain the maintenance of diversity at the plot scale.

\smallskip

Our goal in this study was not to identify what specific mechanisms could be added to the neutral dynamics to explain the maintainence of observed distributions of species abundances.  However, there are several likely ways to resolve this mismatch, and our analysis now opens up the possibility of quantifying what kinds of additional ecological mechanisms provide the best explanation. Very generally, the skew towards rare species in the empirical data can be explained by the presence of stabilizing mechanisms at the local scale. Stabilization can arise from density-dependent interactions, perhaps in turn driven by plant-soil feedbacks~\cite{Mangan2010,Comita2010}, which act to reduce both local dominance and extirpation.  An alternative is that neutral models can still explain the presence of these rare species, but that we need to consider so-called ``fat-tailed'' dispersal, where the probability of dispersing a given distance from a parent tree drops off relatively slowly with distance~\cite{Chave2002b,Rosindell2009,nathan2006long}. Our results raise a challenge to either of these explanations for rare diversity in tropical forests. For example, if plant-soil feedbacks  explain this combination of patterns, can we quantify exactly how strong and at what spatial scales these mechanisms must act?  Similarly, can we quantify exactly what type of long-distance dispersal, if any, can explain the same patterns?  Building on the development of this spatial model to include more general processes will provide a sharp, quantitative test of whether a given proposed mechanism is consistent with observations.

\smallskip

Neutrality has perhaps been tested more than any other single theory of biodiversity.  This scrutiny has ranged across decadal fluctuations~\cite{chisholm2014b,fung2016reproducing,fung2016species} and evolutionary timescales~\cite{Nee2005,chisholm2014ages,Wang2013, odwyer2015phylo,houchmandzadeh2017neutral}, and across taxonomic groups and environments~\cite{Volkov2003,muneepeerakul2008,oficteru2010,woodcock2007,odwyer2015phylo}.  In this manuscript, we show that in terms of the patterns where it has seen greatest success,  species abundance distributions, we are seeing discrepancies between the theoretical predictions and observed data. On the other hand, the precise formulation of the neutral theory is exactly what makes it possible to perform these quantitative tests. While the presence of species differences and local niche structure has also been extensively tested, it has rarely been possible to translate the existence of these mechanisms into quantitative predictions for biogeographical patterns, like the distribution of species abundances as a function of spatial scale. The approach we have taken and discrepancies we have identified may therefore serve to motivate new, and more accurate, models of biodiversity, taking us a step closer to identifying precisely what mechanisms do and do not matter for the prediction of biodiversity patterns~\cite{hubbell2005neutral,mcgill2010towards}.






\section*{Acknowledgments}
J.O.D. acknowledges the Simons Foundation Grant \#376199, McDonnell Foundation Grant \#220020439, and Templeton World Charity Foundation Grant \#TWCF0079/AB47. This work was supported by the Natural Environment Research Council grant number NE/H007458/1.  We thank Ryan Chisholm for comments on an earlier draft, and gratefully acknowledge James Rosindell for permission to use simulated neutral model data generated in collaboration with S.J.C. as a means to test our analytical approximations. The BCI forest dynamics research project was made possible by NSF grants to S. P. Hubbell: DEB \#0640386, DEB \#0425651, DEB \#0346488, DEB \#0129874, DEB \#00753102, DEB \#9909347, DEB \#9615226, DEB \#9405933, DEB \#9221033, DEB \#-9100058, DEB \#8906869, DEB \#8605042, DEB \#8206992, DEB \#7922197, support from CTFS, the Smithsonian Tropical Research Institute, the John D. and Catherine T. MacArthur Foundation, the Mellon Foundation, the Small World Institute Fund, and numerous private individuals, and through the hard work of over 100 people from 10 countries over the past two decades. The plot project is part the Center for Tropical Forest Science, a global network of large-scale demographic tree plots.

\pagebreak

{\Large
\textbf{Supplementary Materials: Methods and Derivations}
}

\appendix
\section{Backward Equation Derivation}

We consider the function $G(z,t)=\sum_{k=0}^{\infty}z^kP(k,t)$, which generates the probability $P(k,t)$ that there are $k$ individuals at time $t$, under the branching process with mortality rate $b$, birth rate $b-\nu$, and with some initial condition at time $t=0$.  Let's now consider the 1-d spatial problem, starting with the problem defined on a lattice.  So now there is birth, death, and hopping to a neighboring lattice site. We define the probability $P(k,1,L,x,t)$, the probability that a single individual at lattice position $x$, a time $t$ in the past, will have $k$ descendants within a region $L$ at the present time.  This then satisfies:
\begin{align}
P(k,1,L,x,t+\Delta t) & = \left(1-(b-\nu)\Delta t - b\Delta t - 2\tilde{D}\Delta t\right)P(k,1,L,x,t) \nn\\&+(b-\nu)\Delta t\sum_m P(k-m,1,L,x,t) P(m,1,L,x,t) +b\Delta t\delta_{k,0}\nn\\
&+\tilde{D}\Delta t \left(P(k,1,L,x+\Delta,t) +P(k,1,L,x-\Delta,t) \right)
\end{align}
where $\tilde{D}$ is the hopping rate and $\Delta$ is the lattice spacing. Taking the limit of $\Delta t$ going to zero gives:
\begin{align}
\frac{\partial P(k,1,L,x,t)}{\partial t} & = -\left(2b-\nu + 2\tilde{D}\right)P(k,1,L,x,t) \nn\\&+(b-\nu)\sum_m P(k-m,1,L,x,t) P(m,1,L,x,t) +b\delta_{k,0}\nn\\
&+\tilde{D}\left(P(k,1,L,x+\Delta,t) +P(k,1,L,x-\Delta,t) \right).
\end{align}
Then, taking the limit of lattice spacing going to zero and defining $D=\Delta^2 \tilde{D}$,
\begin{align}
\frac{\partial P(k,1,L,x,t)}{\partial t} & = -\left(2b-\nu + 2\frac{D}{\Delta^2}\right)P(k,1,L,x,t) \nn\\&+(b-\nu)\sum_m P(k-m,1,L,x,t) P(m,1,L,x,t) +b\delta_{k,0}\nn\\
&+\frac{D}{\Delta^2}\left(P(k,1,L,x,t) +P(k,1,L,x,t) \right)+D\frac{\partial^2 P(k,1,L,x,t)}{\partial x^2}\nn\\
& = -\left(2b-\nu\right)P(k,1,L,x,t)+(b-\nu)\sum_m P(k-m,1,L,x,t) P(m,1,L,x,t) \nn\\&+b\delta_{k,0}
+D\frac{\partial^2 P(k,1,L,x,t)}{\partial x^2}
\end{align}
Finally, we are going to define the generating function of this quantity as:
\begin{align}
G(z,x,t,L) = \sum z^kP(k,1,L,x,t)
\end{align}
so that
\begin{align}
\frac{\partial{G}}{\partial t}&=\left(-2b+\nu\right)G +(b-\nu)G^2 +b +D\frac{\partial^2 G}{\partial x^2}\label{eq:backsi}.
\end{align}
Note that the boundary condition is that $G(z,x,0,L)$ is $=z$ if $x$ is within the sampling region defined by $L$ (e.g. a line segment of length $L$ in the one dimensional problem, and an area of whatever geometry in the 2d case).

\section{Solutions}

\subsection{Species Area Curve}

We now consider an approximation method to find solutions of Eq.~\eqref{eq:backsi}.  First we define:
\begin{equation}
\phi(x,t,L) = 1-G(x,0,t,L)
\end{equation}
so that $\phi$ satisfies:
\begin{equation}
\frac{\partial{\phi}}{\partial t}=-\nu\phi -(b-\nu)\phi^2 +D\frac{\partial^2 \phi}{\partial x^2}\label{eq:backphi}.
\end{equation}
with an initial condition $\phi(x,t=0,L) = R(x,L)$, where $R(x,L)$ is a rectangular function, equal to zero for $x<-L/2$ and $x>L/2$, and equal to one for $-L/2 <x <L/2$.  

The function $\phi(x,t)$ is the probability that an individual appearing in a speciation event at a time $t$ in the past, and at location $x$, will have one or more descendents in the focal region between $-L/2$ and $+L/2$ in the present day.  In order to derive the Species-Area relationship from this probability distribution (where `area' indicates the one-dimensional length of the focal region, $L$), we need to integrate over all speciation events, which in the neutral model occur at a rate $\nu\rho$ per unit time per unit area, where $\rho$ is the equilibrium density of individuals in space. Hence, our goal is to derive a solution for:
\begin{equation}
S(L) = \nu\rho\int_{-\infty}^{\infty} dx \int_0^\infty dt\ \phi(x,t,L).
\end{equation}
Due to the nonlinearity in Eq.~\eqref{eq:backphi}, solving for $\phi(x,t,L)$ exactly does not seem tractable.  But the initial and final conditions for $\phi$ suggest a linear approximation, if we treat $\phi^2(x,t,L) \simeq \phi(x,t,L)R(x,L)$.  While true at $t=0$, and true at late times when $\phi\rightarrow 0$, this does not hold for general $t$, and with this approximation for $S(L)$ we would underestimate the number of species at large values of $L$. The problem is clear---at intermediate times, $\phi(x,t,L)$ will be non-zero outside of the focal region, and will interpolate between one and zero in side the focal region.

We therefore handle this discrepancy by approximating $\phi^2(x,t,L)$ as $\phi^2(x,t,L) =\phi(x,t,L)$ while $x$ is within the focal region, and outside of the focal region we set $\phi^2(x,t,L) \propto\phi(x,t,L)$ with a (we expect small) constant of proportionality to be determined.  This leads to an equation of the form:
\begin{equation}
\frac{\partial{\phi}}{\partial t}=-b\nuff\phi -b(1-\nuff)\phi R(x,L) +D\frac{\partial^2 \phi}{\partial x^2}\label{eq:backphiapp}.
\end{equation}
So in fact, we have two equations to solve, as the approximation we have used leads to different equations inside and outside of the focal region defined by $L$:
\begin{align}
\frac{\partial{\phin}}{\partial t} & =-b\phin+D\frac{\partial^2 \phin}{\partial x^2}\nn\\
\frac{\partial{\phout}}{\partial t} & =-b\nuff\phout +D\frac{\partial^2 \phout}{\partial x^2}
\end{align}
We can now integrate over time, before solving these equations as a function of space, to obtain:
\begin{align}
-b\nuff\rho & =-b\Phin+D\frac{\partial^2 \Phin}{\partial x^2}\nn\\
0 & =-b\nuff\Phout +D\frac{\partial^2 \Phout}{\partial x^2}\label{eq:inout}
\end{align}
where $\Phin(x,L) = \nuff\rho\int_0^\infty dt\ \phin(x,t,L)$  and $\Phout(x,L) =\nuff\rho \int_0^\infty dt\ \phout(x,t,L)$, respectively.  Note that we have also introduced a new, effective rate of introduction of new species per unit space and time, $b\nuff\rho$, instead of $\nu\rho$, for consistency at small values of L with the term $\nuff\phout$ in Eq.~\eqref{eq:inout}. It may seem like we have introduced a free parameter or parameters by allowing for $\nuff$, but in fact this effective rate is fixed by the large scale behavior, i.e. as $L\rightarrow\infty$.  In this limit,
\begin{equation}
-\nuff\rho  =-\Phin
\end{equation}
which leads to a solution
\begin{align}
S(L\rightarrow\infty) = \rho L \nuff.
\end{align}
So in order to match the standard neutral result for a well-mixed community, at large scales we have that
\begin{equation}
\nuff(\nu) = -\frac{\nu}{b-\nu}\log(\nu/b)
\end{equation}
with no free parameters.

\smallskip

We can solve the pair of equations~\eqref{eq:inout} by imposing that there is no singular behavior, that $\Phout$ asymptotes to zero for large $x$, and that at the boundaries $\pm L/2$ both $\Phout$ and $\Phin$ and their first derivatives match.  The result is:
\begin{align}
\Phin(x,L) & = \nuff\rho\lt(1 -  \frac{\cosh(mx)}{{\frac{1}{\sqrt{\nuff}}\sinh(mL/2)+\cosh(mL/2)}}\rt)
\nn\\
\Phout(x,L) & =  \rho\frac{e^{-m|x-L/2|}\sinh(mL/2)}{\frac{1}{\sqrt{\nuff}}\sinh(mL/2)+\cosh(mL/2)}
\end{align}
where we have defined the inverse length-scale $m = \sqrt{b/D}$, also denoted by $m=1/\sigma$ in the main text, that is associated with the diffusion or dispersal process driving the spatial distribution of these neutral organisms.  We now integrate over space to obtain our approximate prediction for the one-dimensional Species-Area relationship.
\begin{align}
S = \nuff\rho\lt[L+\frac{2(b/\nuff-1)}{m}\frac{\tanh(mL/2)}{\frac{1}{\sqrt{\nuff}}\tanh(mL/2)+1}\rt]
\end{align}
In the following subsections we will show that despite our approximation, this provides an extremely accurate prediction of the relationship across a broad range of areas.

We now provide the corresponding result in two spatial dimensions.  We apply exactly the same approximation, but where now we interpret $R(x,y,L)$ as the `top-hat' function, which is equal to one inside a circular region of radius $L$, and is equal to zero outside.  We then solve for functions inside and outside of this circular region:
\begin{align}
-b\nuff\rho & =-b\Phin+D\nabla^2\Phin \nn\\
0 & =-b\nuff\Phout +D\nabla^2\Phout\label{eq:inout2d}
\end{align}
This leads to solutions which depend only on a radial coordinate, $r$ (distance from the origin), and not on the corresponding polar coordinate:
\begin{align}
\Phin(r,L) & = \nuff\rho\lt(1 -  \frac{I_0(mr)}{\frac{1}{\sqrt{\nuff}}I_1(mL)\frac{K_0\lt(mL\sqrt{\nuff}\rt)}{K_1\lt(mL\sqrt{\nuff}\rt)}+I_0(mL)}\rt)
\nn\\
\Phout(r,L) & =\rho\frac{\sqrt{\nuff}\ I_1(mL)\frac{K_0\lt(mr\sqrt{\nuff}\rt)}{K_1\lt(mL\sqrt{\nuff}\rt)}}{\frac{1}{\sqrt{\nuff}}I_1(mL)\frac{K_0\lt(mL\sqrt{\nuff}\rt)}{K_1\lt(mL\sqrt{\nuff}\rt)}+I_0(mL)}
\end{align}
and integrating over all space we find:
\begin{align}
S(\textrm{radius} = L) = \nuff\rho\lt[\pi L^2+\frac{2\pi L(1/\nuff-1)}{m}\frac{I_1(mL)}{\frac{1}{\sqrt{\nuff}}I_1(mL)\frac{K_0(m\sqrt{\nuff}L)}{K_1(m\sqrt{\nuff}L)}+I_0(mL)}\rt]
\end{align}
where again $\nuff= -\frac{\nu}{1-\nu}\log(\nu/b)$.  In terms of sample area $A = \pi L^2$ we can rewrite this as:
\begin{align}
S(A) = \nuff\rho\lt[A+\frac{2\sqrt{\pi}\sqrt{A} (1/\nuff-1)}{m}\frac{I_1(\frac{m\sqrt{A}}{\sqrt{\pi}})}{\frac{1}{\sqrt{\nuff}}I_1(\frac{m\sqrt{A}}{\sqrt{\pi}})\frac{K_0(m\sqrt{A\nuff/\pi})}{K_1(m\sqrt{A\nuff/\pi})}+I_0(\frac{m\sqrt{A}}{\sqrt{\pi}})}\rt]
\end{align}
In the main text we replaced the inverse length-scale $m$ with a length-scale $\sigma=1/m$, but both can be related directly to the parameter $D$ we introduced in the formulation of this problem.

\subsection{Species Abundance Distribution}

We now consider the same kind of approximation method to find solutions of Eq.~\eqref{eq:backsi}, but instead of considering just total species richness, we define (again first considering the one-dimensional case):
\begin{equation}
\phi(x,z,t,L) = 1-G(x,z,t,L)
\end{equation}
so that $\phi$ again satisfies:
\begin{equation}
\frac{\partial{\phi}}{\partial t}=-\nu\phi -(b-\nu)\phi^2 +D\frac{\partial^2 \phi}{\partial x^2}.\label{backphi}
\end{equation}
with an initial condition $\phi(x,z,t=0,L) = (1-z)R(x,L)$, where $R(x,L)$ is the same rectangular function as above.  To obtain the Species Abundance Distribution, we need to integrate over all speciation events, which in the neutral model occur at a rate $\nu\rho$ per unit time per unit area, where $\rho$ is the equilibrium density of individuals in space. Hence, our goal is to derive a solution for:
\begin{equation}
\Psi(z,L) =S(L)- \nu\rho\int_{-\infty}^{\infty} dx \int_0^\infty dt\ \phi(x,t,L).\label{Psiint}
\end{equation}
where using this definition, $\Psi(z,L)$ is related to the Species Abundance Distribution in a sample taken from the region of size $L$ by
\begin{equation}
\Psi(z,L) = \sum_{k=1}^{\infty}S(k,L)  z^k.
\end{equation}

\smallskip

We now extend our previous approximation for the species area curve.  We follow the same principle to set $\phi^2(x,z,t,L) \simeq (1-z)\phi(x,z,t,L)$ when $x$ is within the focal region, while outside of the focal region, we set $\phi^2(x,z,t,L) =g(z)\phi(x,z,t,L)$ for a function $g(z)$ to be determined by the requirement that we match the known behavior at large values of $L$.  This leads to an equation of the form:
\begin{equation}
\frac{\partial{\phi}}{\partial t}=-g(z)\phi -(b(1-z)+\nu z-g(z))\phi R(x,L) +D\frac{\partial^2 \phi}{\partial x^2}\label{eq:backphiapp}.
\end{equation}
Again, we have two equations to solve, as this approximation leads to different equations inside and outside of the focal region defined by $L$:
\begin{align}
\frac{\partial{\phin}}{\partial t} & =-(b(1-z)+\nu z)\phin+D\frac{\partial^2 \phin}{\partial x^2}\nn\\
\frac{\partial{\phout}}{\partial t} & =-bg(z)\phout +D\frac{\partial^2 \phout}{\partial x^2}
\end{align}
We can now integrate over time, before solving these equations as a function of space, to obtain:
\begin{align}
-bg(z)(1-z)\rho & =-(b(1-z)+\nu z)\Phin+D\frac{\partial^2 \Phin}{\partial x^2}\nn\\
0 & =-bg(z)\Phout +D\frac{\partial^2 \Phout}{\partial x^2}\label{eq:inoutsad}
\end{align}
where $\Phin(x,z,L) = bg(z)\rho\int_0^\infty dt\ \phin(x,z,t,L)$  and $\Phout(x,z,L)=bg(z)\rho \int_0^\infty dt\ \phout(x,z,t,L)$, respectively.  Note that we have also introduced a new, effective rate of introduction of new species per unit space and time, $bg(z)\rho$, instead of $\nu\rho$, for consistency at small values of L with the term $bg(z)\phout$ in Eq.~\eqref{eq:inoutsad}. The function $g(z)$ is then fixed by the large scale behavior. In this limit,
\begin{equation}
bg(z)	(1-z)\rho  =(b(1-z)+\nu z)\Phin
\end{equation}
which leads to a large-scale solution
\begin{align}
\Psi(z,L\rightarrow\infty) = \rho L \frac{\nu}{b-\nu}\log(b/\nu) -\rho L \frac{g(z)(1-z)}{(1-z+\nu z/b)}.
\end{align}
So in order to match the standard neutral result for a well-mixed, non-zero-sum community~\cite{Volkov2003}, which is
\begin{align}
\Psi(z,L) & = \sum_{k=1}^{\infty} S_{\textrm{nzs}}(k) z^k\nn\\
& =  \sum_{k=1}^{\infty}\frac{\nu\rho L}{b-\nu}\lt(\frac{b-\nu}{b}\rt)^k z^k \nn\\
& = -\frac{\nu\rho L}{b-\nu} \log \lt(1-\frac{b-\nu}{b}z\rt)
\end{align}
we need to set
\begin{align}
g(z) =\frac{1-z+\nu z/b}{1-z} \frac{\nu}{b-\nu}\log\lt(\frac{b-(b-\nu)z}{\nu}\rt).
\end{align}
We note also that $g(0) = \nuff$, and so this approximation simply reduces to the approximation we used to derive the Species-area Curve above; our solution above for $\phi(z,t,L)$ is equal to $\phi(x,z=0,t,L)$ here, so that our approximation for the Species-area curve satisfies these same equations but with $z=0$ (as it should do).

\smallskip

The result of solving this pair of equations is essentially the same as above. The dependence on $z$ of the parameter does not affect the solution as a function of $x$, it just changes the parametrization in that solution:
\begin{align}
\Phin(x,z,L) & = f(z)\rho\lt(1 -  \frac{\cosh(mx\sqrt{h(z)})}{{\sqrt{\frac{1-z}{f(z)}}\sinh(mL\sqrt{h(z)}/2)+\cosh(mL\sqrt{h(z)}/2)}}\rt)
\nn\\
\Phout(x,z,L) & = \rho\sqrt{f(z)(1-z)} \frac{e^{-\sqrt{g(z)}m|x-L/2|}\sinh(mL\sqrt{h(z)}/2)}{\sqrt{\frac{1-z}{f(z)}}\sinh(mL\sqrt{h(z)}/2)+\cosh(mL\sqrt{h(z)}/2)}
\end{align}
where again $m = \sqrt{b/D}=1/\sigma$ in the main text, and for ease of notation we introduce
\begin{align}
h(z) = \lt((1-z)+\frac{\nu}{b} z\rt)\nn\\
f(z)= \frac{\nu}{b-\nu}\log\lt(\frac{b-(b-\nu)z}{\nu}\rt)
\end{align}
as in the main text, and such that $g(z)$, $f(z)$ and $h(z)$ are related by $g(z)=h(z)f(z)/(1-z)$. We now integrate over space to obtain our approximate prediction for the one-dimensional Species Abundance Distribution.
\begin{align}
\Psi_{1d}(z,L) = S(L) - \rho f(z) L + \frac{\frac{2\rho}{m\sqrt{h(z)}}(f(z)-(1-z))\sinh(mL\sqrt{h(z)}/2)}{\sqrt{\frac{1-z}{f(z)}}\sinh(mL\sqrt{h(z)}/2)+\cosh(mL\sqrt{h(z)}/2)}
\end{align}
The same approach in 2 spatial dimensions for a circular region of radius $L$ and area $A=\pi L^2$ leads to:
\begin{align}
\Phin(r,L) & =f(z)\rho\lt(1 -  \frac{I_0(mr\sqrt{h(z)})}{\sqrt{\frac{1-z}{f(z)}}I_1(mL\sqrt{h(z)})\frac{K_0\lt(mL\sqrt{g(z)}\rt)}{K_1\lt(mL\sqrt{g(z)}\rt)}+I_0(mL\sqrt{h(z)})}\rt)
\nn\\
\Phout(r,L) & =\rho\frac{\sqrt{f(z)(1-z)}\ I_1(mL\sqrt{h(z)})\frac{K_0\lt(mr\sqrt{g(z)}\rt)}{K_1\lt(mL\sqrt{g(z)}\rt)}}{\sqrt{\frac{1-z}{f(z)}}I_1(mL\sqrt{h(z)})\frac{K_0\lt(mL\sqrt{g(z)}\rt)}{K_1\lt(mL\sqrt{g(z)}\rt)}+I_0(mL\sqrt{h(z)})}
\end{align}
and
\begin{align}
\Psi_{2d}(z,A) & = S(A)-\rho f(z) A 
 +\frac{ \frac{2(f(z)-(1-z))}{\sqrt{h(z)}}\rho\sqrt{A\pi\sigma^2}\ I_1\lt(\sqrt{\frac{h(z)A}{\pi\sigma^2}}\rt)}{I_0\lt(\sqrt{\frac{Ah(z)}{\pi\sigma^2}}\rt)+\sqrt{\frac{1-z}{f(z)}}\frac{K_0\lt(\sqrt{\frac{Ag(z)}{\pi\sigma^2}}\rt)}{K_1\lt(\sqrt{\frac{Ag(z))}{\pi\sigma^2}}\rt)}I_1\lt(\sqrt{\frac{Ah(z)}{\pi\sigma^2}}\rt)}.\label{Psi2d}
\end{align}

\subsection{Universal behaviour as $\nu\to0$}
Here, we show that the species abundance distributions given by our
model exhibit the same universality property found in simulations
by Rosindell and Cornell \cite{rosindell2013universal} i.e. that the species abundance
distributions form a family of curves, parametrised by the single
paramemeter $\frac{A\nu}{b\sigma^{2}}$. 

First, we show that the exact solution to the backward equation for
$\phi$ has this scaling property. If we define
\begin{align*}
Q & =\frac{b-\nu}{\nu}q\\
T & =\nu t\\
X & =x\sqrt{\frac{\nu}{D}},
\end{align*}
then eqn. (\eqref{eq:backphi}) becomes
\[
\frac{\partial Q}{\partial T}=-Q-Q^{2}+\frac{\partial^{2}Q}{\partial X^{2}},
\]
and the initial condition becomes $Q(X,T=0)=\frac{(b-\nu)(1-z)}{\nu}R(X,L\sqrt{\frac{\nu}{D}}$).
Therefore, $Q$ only depends on the parameters through the combinations
\begin{align*}
Z & =\frac{(b-\nu)(1-z)}{\nu}\\
Y & =\frac{A\nu}{D}=\frac{A\nu}{b\sigma^{2}}
\end{align*}
i.e. $Q=\tilde{Q}(X,T,Z,Y)$ for some function $\tilde{Q}$. The generating
function for the abundance distribution in 2D is then given by
\begin{align}
\Psi & =S(A)-\nu\rho\int_{-\infty}^{\infty}\int_{-\infty}^{\infty}d^{2}x\int_{0}^{\infty}dt\,\phi(x,t)\nonumber\\
 & =S(A)-\frac{\rho b\sigma^{2}}{b-\nu}\int_{-\infty}^{\infty}\int_{-\infty}^{\infty}d^{2}X\int_{0}^{\infty}dT\,\tilde{Q}(X,T,Z,Y)\nonumber\\
 & =S(A)-\frac{\rho b\sigma^{2}}{b-\nu}\tilde{\Psi}(Z,Y)\nonumber\\
 & \to S(A)-\rho \sigma^2\tilde{\Psi}(Z,Y)+O(\nu)\label{Psiscale}
\end{align}
for some function $\tilde{\Psi}$.

While we do not have an expression for the exact solution $\Psi$,
we can verify that our approximate solution $\Psi_{2d}$ has the same
scaling behaviour. Substituting $1-z=\frac{Z\nu}{b-\nu}$, $A=\frac{Yb\sigma^{2}}{\nu}$
we get
\begin{align*}
h(z) & =\frac{\nu}{b}(1+Z)+O(\nu^{2})\\
f(z) & =\frac{\nu}{b}\log\left(1+Z\right)+O(\nu^{2})\\
g(z) & =\frac{\nu}{b}(1+Z)\log\left(1+Z\right)+O(\nu^{2})\\
\frac{f(z)-(1-z)}{\sqrt{h(z)}}\sqrt{A\pi\sigma^{2}} & =(\log\left(1+Z\right)-Z)\sigma^{2}\sqrt{\pi Y}+O(\nu)\\
\frac{h(z)A}{\pi\sigma^{2}} & =\frac{Y(1+Z)}{\pi}+O(\nu)\\
\frac{g(z)A}{\pi\sigma^{2}} & =\frac{Y(1+Z)\log(1+Z)}{\pi}+O(\nu)
\end{align*}
so eqn. (\eqref{Psi2d}) becomes
\[
\Psi_{2d}=S(A)-\rho\sigma^{2}Y\log(1+Z)+2\rho\sigma^{2}\frac{(\log\left(1+Z\right)-Z)\sqrt{\pi Y}I_{1}\left(\sqrt{\frac{Y(1+Z)}{\pi}}\right)}{I_{0}\left(\sqrt{\frac{Y(1+Z)}{\pi}}\right)+\sqrt{\frac{Z}{\log(1+Z)}}\frac{K_{0}\left(\sqrt{\frac{Y(1+Z)\log(1+Z)}{\pi}}\right)}{K_{1}\left(\sqrt{\frac{Y(1+Z)\log(1+Z)}{\pi}}\right)}I_{1}\left(\sqrt{\frac{Y(1+Z)}{\pi}}\right)}+O(\nu)
\]
which is of the same form as eqn. (\eqref{Psiscale}).

We have thus shown that the generating function of the abundance distribution
is a function of two parameter combinations only. We will now show
that this is equivalent to the observation that the species abundance
distribution is a one-parameter family of curves \cite{rosindell2013universal}:
\[
S(k,A)=\nu\tilde{S}\left(\nu k,\frac{A\nu}{b\sigma^{2}}\right).
\]
(note that the expression in ref. \cite{rosindell2013universal} describes the scaling of logarithmic
Preston classes of abundance, and hence is missing the prefactor $\nu$).
Note ref. \cite{rosindell2013universal} used a dispersal kernel with length scale
$L$ rather than the Brownian motion used in the present study, but at
the large spatial scales of interest to us a jump process with an
exponentially bounded kernel is equivalent to a random walk with $\sigma=L$.
To see how this is related to our scaling expression for $\Psi(z,A)$,
we write
\begin{align*}
\Psi(z,A) & =\sum_{k=1}^{\infty}z^{k}S(k,A)\\
 & \approx\int_{1}^{\infty}e^{k\log z}\nu\tilde{S}\left(\nu k,\frac{A\nu}{b\sigma^{2}}\right)dk\\
 & =\int_{\nu}^{\infty}e^{m\frac{\log z}{\nu}}\tilde{S}\left(m,Y\right)dm,
\end{align*}
where $m=\nu k$. We need to proceed with caution in case $\tilde{S}$
has a non-integrable singlarity in its first argument. Abundance in
Preston classes of low order appears from Fig 2 in the paper to approach
a finite limit, so we assume that $\tilde{S}\sim\frac{s\left(Y\right)}{\nu k}$
at small $(\nu k)$. Without loss of generality, we write
\begin{align*}
\Psi(z,A) & =\int_{\nu}^{\infty}\left(\frac{1}{m}s\left(Y\right)e^{-m}+e^{m\frac{\log z}{\nu}}u\left(m,Y\right)-\frac{1}{m}s\left(Y\right)e^{-m}\right)dm\\
 & =s\left(Y\right)\mbox{Ei}(\nu)+f\left(\frac{\log z}{\nu},Y\right)+O(\nu)\\
 & =s\left(Y\right)\mbox{Ei}(\nu)+f\left(\frac{\log\left(1-\frac{Z\nu}{b-\nu}\right)}{\nu},Y\right)+O(\nu)\\
 & =s\left(Y\right)\mbox{Ei}(\nu)+f\left(Z,Y\right)+O(\nu),
\end{align*}
where $\mbox{Ei}(x)=\int_{x}^{\infty}\frac{\exp(-y)}{y}dy$ is the
exponential integral and $f$ is a (finite) function of two arguments.
The component of this expression that depends on $Z$ takes the same
scaling form as found above for the backward equation model described
above. The term that is independent of $Z$ does not contribute to
any particular $S(k,A)$, but does contribute to the total species
richness, and indeed we can identify
\begin{align*}
S(A) & =\Psi(1,A)\\
 & =s\left(Y\right)\mbox{Ei}(\nu)+f\left(0,Y\right)
\end{align*}

\section{Biological Interpretation of the Approximation Method}

By approximating the non-linear term in the defining backward Equation~\eqref{eq:backphi} by a heterogeneous linear term, we found the pair of equations~\eqref{eq:inout} to solve for the species area relationship:
\begin{align}
-b\nuff\rho & =-b\Phin+D\frac{\partial^2 \Phin}{\partial x^2}\nn\\
0 & =-b\nuff\Phout +D\frac{\partial^2 \Phout}{\partial x^2}\nn.
\end{align}
(For simplicity we work in one spatial dimension but the interpretations are identical in $2d$.)  We could equally well interpret these not just as an approximation to Eq.~\eqref{eq:backphi}, but as a biological model in their own right.  If we do so, can we reinterpret these equations and understand biologically why this linear approximation works?  Eqs.~\eqref{eq:inout} constitute a system where there is only mortality (driving loss of species from the focal, sample area), dispersal, and input from speciation. This might be expected, since species are only removed from the focal region when there is a mortality event, and only added when there is a speciation event landing in the focal region, or dispersal in from outside.  However, in these equations the effective rates of species loss are different for species which originated outside the focal region (rate $\nuff$), versus those that originated inside the focal region (rate $b$), and that is what we must explain.

\smallskip

Remembering that in the original derivation above of Eq.~\eqref{eq:backphi}, the per capita birth rate was $b-\nu$ and mortality rate was $b$, this interpretation of rate of species loss in the equation for $\Phin$ becomes clear: species are lost from the focal region at the rate at which a single individual dies. I.e. we are approximating that species which originate within $-L/2 <x<L/2$, will only \textit{not} be found at time $t$ in this sample region if it goes extinct, and this rate of loss is approximated by the rate of loss $b$ of a single individual. For species outside the sample region, the rate of loss from mortality is $b\nuff=\frac{b\nu}{b-\nu}\log(b/\nu)$.  What is this number?  In fact, it is equal to $b/\la n\ra$, where $\la n\ra$ is the expected population size of an extant species (i.e. total number of individuals divided by total number of extant species).  On average, a species originating outside the focal region at any point in the past, is lost from the focal region at a effective rate, $b/\la n\ra$. 

\section{Comparison with Field Theory/Forward-in-time Equations}

In an earlier paper, one of the authors derived a forward-in-time approach to these same spatial neutral models~\cite{Odwyer2010}. In that approach, we also began with the case of a spatially-discrete landscape.  Here we will recap the basic features and approximation we made in that paper, and where they break down relative to our current approach. We will also work directly with individuals that diffuse across the landscape as we have in this paper 

We first describe the state of the discrete system using the probability distribution $P(\dots,n_i\dots,t)$ that there are $n_i$ individuals at each spatial location, $i$, at time $t$, belonging to a focal species. Individuals in this spatially-discrete model die with a per capita mortality rate, $d$,  produce new offspring at a per capita birth rate, $b$, and may transfer to nearest neighbour cells at a rate $\tilde{D}$. In addition, there is a speciation process modeled as immigration from outside the system at a rate $\tilde{k}$ from $0$ to $1$ individual:
\begin{align}
 \frac{\partial P(\{n_i\},t)}{\partial t} & = d\sum_{i}
(n_i+1)P(\dots,n_i+1,\dots,t)
-d\sum_{i}n_iP(\dots,n_i,\dots,t) \nn\\&+b\sum_{i}(n_i-1)P(\dots,n_i-1,\dots,t) - b\sum_{i}n_iP(\dots,n_i,\dots,t)\nn\\ & + \tilde{D}\sum_i\sum_{\{e\}}\lt[(n_i+1)P(\dots,n_i+1,n_e-1,t)-n_iP(\dots,n_i,n_e,\dots)\rt]\nn\\
& + \tilde{k}\sum_i  \lt(\delta_{n_i 1}\prod_{j\neq i}\delta_{n_j 0}\rt)  -\tilde{k} \sum_i  \lt(\prod_{j}\delta_{n_j 0} \rt)\label{eq:discrete1}
\end{align}
We could also remove this last term, introducing speciation, and thus allow each species to reach permanent extinction. We would then sum the contributions to the present day state from all species that originated at some point in the past, assuming a uniform speciation rates across time and space. Before taking the limit of continuous space, we rewrite the dynamics of our discrete community in terms of a moment generating function. This generating function is defined by a sum over all spatial configurations of individuals:
\begin{equation}
Z(\dots,h_i,\dots,t)=\sum_{\{n_{k}\}}P(\dots,n_{i},\dots,t)e^{\sum_jh_{j}n_{j}}\label{eq:discgen}.
\end{equation}
Rewriting Eq.\eqref{eq:discrete1} in terms of this generating function, we find a new defining equation: 
\begin{align}
\frac{\partial Z}{\partial t} & = d\sum_{i=-\infty}^{\infty}
\frac{\partial Z}{\partial h_i}\lt(e^{-h_i}-1\rt) +b\sum_{i=-\infty}^{\infty}
\frac{\partial Z}{\partial h_i}\lt(e^{h_i}-1\rt) +\tilde{D}\sum_{i=-\infty}^{\infty}\sum_{\{e\}}
\frac{\partial Z}{\partial h_i}\lt(e^{h_e-h_i}-1\rt) +\tilde{k} \sum_{i=-\infty}^{\infty}\lt(e^{h_i}-1\rt).
\label{eq:discrete2}
\end{align}

\subsection{Taking a Continuum Limit}

We denote the lattice spacing by $\Delta$, and define the continuum limit as follows:
\begin{align}
\sum_i & \rightarrow \Delta^{-d} \int d^dx\nn\\
h_i & \rightarrow  H(x)\nn\\
\frac{\partial}{\partial h_i} & \rightarrow \Delta^{d}\frac{\delta}{\delta H(x)}\nn\\
\tilde{D} & \rightarrow \frac{D}{\Delta^2}\nn\\
\tilde{k}& \rightarrow k\Delta^d
\end{align}
Finally, to define the continuum limit for the sum over nearest neighbours, we consider a square, $d$-dimensional lattice:
\begin{align}
\sum_{\{e\}}\lt(e^{h_e-h_i}-1\rt) & \rightarrow \sum_{k=1}^d \lt(\exp\lt({\Delta\frac{\partial H}{\partial x_k}+\frac{\Delta^2}{2}\frac{\partial^2 H}{\partial x_k^2}+\dots}\rt) +\exp\lt({-\Delta\frac{\partial H}{\partial x_k}+\frac{\Delta^2}{2}\frac{\partial^2 H}{\partial x_k^2}+\dots}\rt)-2\rt)\nn\\
&= \Delta^2\lt(\nabla^2 H(x) +(\nabla H(x))^2\rt)+O(\Delta^3)
\end{align}
With these identifications, the multivariate generating function Eq.~\eqref{eq:discgen} becomes a functional of the source $H(x)$:
\begin{equation}
Z[H(x),t] = \lt\langle e^{\int dx H(x)n(x)}\rt\rangle
\end{equation}
where $\langle n(x)\rangle$, $\langle n(x_1) n(x_2)\rangle$ etc are expectation values of the number densities and correlations of individuals as a function of spatial location (express this better). This generating functional satisfies the continuum limit of Eq.~\eqref{eq:discrete2}, the following functional differential equation:
\begin{align}
\frac{\partial Z}{\partial t} & = d \int d^dx
\frac{\delta Z}{\delta H(x)}\lt(e^{-H(x)}-1\rt) +b \int d^dx
\frac{\delta Z}{\delta H(x)}\lt(e^{H(x)}-1\rt) \nn\\ 
& +D\int d^dx
\frac{\delta Z}{\delta H(x)}\lt(\nabla^2 H(x) +(\nabla H(x))^2\rt) +kZ \int d^dx\lt(e^{H(x)}-1\rt)\nn\\ 
& =\int d^dx \lt(e^{H(x)}-1\rt)\lt(
-de^{-H(x)}\frac{\delta Z}{\delta H(x)} +b
\frac{\delta Z}{\delta H(x)} +D
\nabla^2\lt(e^{-H(x)}\frac{\delta Z}{\delta H(x)}\rt)  +k \rt).
\end{align}
where we have performed an integration by parts and assumed that the source $H(x)$ vanishes at infinity. Finally, we make a change of variables for the source, 
\begin{equation}
J(x) = e^{H(x)}-1
\end{equation}
so that $\mathcal{Z}[J(x),t] = Z[\log(J(x)+1),t]$ satisfies
\begin{align}
\frac{\partial  \mathcal{Z}}{\partial t}  & = \int d^dx\ J(x)\lt(
(b-d)\frac{\delta \mathcal{Z}}{\delta J(x)} +bJ(x)
\frac{\delta \mathcal{Z}}{\delta J(x)} +D
\nabla^2\frac{\delta \mathcal{Z}}{\delta J(x)}  +k\rt)\label{eq:finalfunctional}.
\end{align}

\subsection{Equal Time Correlation Functions}

The $n$-point spatial correlation functions for this model, taken at equal times, $t$, satisfy a set of partial differential equations. These equations are obtained by expanding $\mathcal{Z}[J(x),t]$ as a functional Taylor series:
\begin{align}
\mathcal{Z}[J(x),t]  & = \int d^dx\ c_1(x,t)J(x) + \frac{1}{2}\int d^dx_1 d^dx_2\ c_2(x_1,x_2,t) J(x_1)J(x_2) + \dots
\end{align}
 The coefficients of this Taylor series are obtained by taking functional derivatives of Eq.\eqref{eq:finalfunctional} with respect to $J$, and then setting $J=0$.  For the first two orders we have:
\begin{align}
\frac{\partial c_1}{\partial t} & =   (b-d) c_1 +D\nabla^2 c_1+k\nn\\
\frac{\partial c_2}{\partial t} & = 2b c_1\delta(x_1-x_2)+D\lt(\nabla_1^2+\nabla_2^2+2(b-d)\rt)c_2(x_1,x_2,t)
\end{align}
Each successive order relies only on solutions for correlation functions of lower order, and so the system of linear partial differential equations can be solved exactly, given a set of initial data.

\subsection{Species Area Relationship}

We now consider the time-independent probability $P(N,L)$ that at late times there are $N$ individuals in a given sample region extending from $-L$ to $+L$.  The generating function of this probability is:
\begin{equation}
\psi(j,L) = \sum_{N=0}^{\infty} P(N,L)(1+j)^N = \lt\la e^{ \log(1+j) \int_L dx\ n(x)} \rt\ra.
\end{equation}
To underline the interpretation: $P(N,L)$ is the (assumed time-independent solution for the) probability distribution that we will find $N$ individuals in the region between $-L$ and $+L$ at late times. The second equality arises because this generating function can be obtained by setting $J(x)=j\textrm{Rect}(x,L)$ in the late-time, time-independent solution for $Z[J,t]$, where $\textrm{Rect}(x,L)$ is the rectangular function in $1$d.  I.e. we will set $J(x)$ zero outside the sample region and equal to $j$ inside. The expected number of species in the sampling region defined by $L$ is proportional to $1-P(0,L)$, and so this is the quantity we are aiming to solve for. If we can solve for this generating function then we have $P(0,L)=\psi(-1,L)$.  This is also known as the empty interval function, e.g.~\cite{Doering1989}.  To find $\psi(-1,L)$, we next define the modified moments which have insertions of $ e^{ \log(1+j)\int_L n(x)}$ compared with the usual moments:
\begin{align}
f_1(x,j,L) & =  \lt.\frac{\delta \mathcal{Z}[J]}{\delta J(x)}\rt|_{J=j\textrm{Rect}(x,L)}\nn\\
& = \frac{1}{1+j\textrm{Rect}(x,L)}\lt\langle n(x) e^{ \log(1+j)\int_L n(x)}\rt\rangle\label{eq:f1def}\\
f_2(x,y,j,L) & = \lt.\frac{\delta^2 \mathcal{Z}[J]}{\delta J(x)\delta J(y)}\rt|_{J=j\textrm{Rect}(x,L)}\nn\\
& = ...
\label{eq:sadgenx}
\end{align}
We note that the reason for using these functions is that:
\begin{align}
\frac{\partial\psi}{\partial j} = \frac{1}{1+j}\lt\langle \int_L dx n(x) e^{ \log(1+j)\int_L n(x)}\rt\rangle = \int_L dx f_1(x,j,L).
\end{align}
and hence if we can solve for $f_1(x,k,L)$ we will have the empty interval function, $P(0,L)$. 

We can obtain differential equations for these modified moments by taking successive functional derivatives of  Eq.\eqref{eq:finalfunctional}:
\begin{align}
\frac{\partial}{\partial t}\frac{\delta \mathcal{Z}[J,t]}{\delta J(x)}  & =D\nabla^2\frac{\delta \mathcal{Z}}{\delta J(x)}+(bJ(x)+b-d)\frac{\delta \mathcal{Z}}{\delta J(x)} + k \nn\\
& +\int dy\ J(y)\frac{\delta}{\delta J(x)}\lt[D\nabla^2\frac{\delta \mathcal{Z}}{\delta J(y)}+(bJ(y)+b-d)\frac{\delta \mathcal{Z}}{\delta J(y)} + k \rt]\nn\\
\frac{\partial}{\partial t}\frac{\delta^2 \mathcal{Z}[J,t]}{\delta J(x)\delta J(y)} & = ...
\end{align}
etc. So setting $J(x)=j\textrm{Rect}(x,L)$ and time derivatives equal to zero in these equations we have a kind of moment hierarchy:
\begin{align}
0 & =D\nabla_x^2f_1(x,j,L)+(bj\textrm{Rect}(x,L)+b-d)f_1(x,j,L) + k \nn\\
& + j\int_L dy\ \lt[D\nabla_y^2f_2(x,y,j,L)+(bj+b-d)f_2(x,y)+b\delta(x-y)f_1(y,j,L)\rt]\\
0 & = D\nabla^2_x f_2(x,y,j,L) +...
\end{align}
So far there is no approximation. In our earlier paper we truncated and solved the first equation in this hierarchy:
\begin{align}
0 & =D\nabla^2f_1(x,j,L)+(bj\textrm{Rect}(x,L)+b-d)f_1(x,j,L) + k
\end{align}
While giving qualitatively accurate description of the shape of the SAR, this earlier approximation becomes quantitatively inaccurate as a speciation rate becomes small. It also fails to give a good description of the Species Abundance Distribution.  

\section{Numerical Inversion of the SAD Generating Function}

We implemented a method described in~\cite{bornemann2011accuracy} for numerical contour integration using Cauchy's theorem. Our annotated and documented code will be freely available on the O'Dwyer lab GitHub repository.







\end{document}